\documentclass[12pt]{article}
\usepackage{cite}
\usepackage{amsmath}
\usepackage[pdftex,unicode,colorlinks,citecolor=blue]{hyperref}
\usepackage{cmap}
\newtheorem{thm}{Theorem}

\newtheorem{lem}{Lemma}
\newtheorem{rem}{Remark}
\newtheorem{deff}{Definition}

\def\baselinestretch{1.5}
\numberwithin{equation}{section}
\begin{document}

\begin{center}
{\Large Characteristic Lie rings, finitely-generated modules and integrability conditions for 2+1 dimensional lattices}
\end{center}

\begin{center}
{Ismagil Habibullin}\footnote{e-mail: habibullinismagil@gmail.com}\\

{Ufa Institute of Mathematics, Russian Academy of Science,\\
Chernyshevskii Str., 112, Ufa, 450077, Russia}
\end{center}

\begin{abstract}
Characteristic Lie rings for Toda type 2+1 dimensional lattices are defined. Some properties of these rings are studied. Infinite sequence of special kind modules are introduced. It is proved that for known integrable lattices these modules are finitely generated. Classification algorithm based on this observation is briefly discussed. 
\end{abstract}

\def\baselinestretch{1.5}


\newpage
\tableofcontents

\section{Introduction}

In the last decade the problem of classification of integrable multidimensional models is intensively studied.
A very powerful approach to the problem based on hydrodynamic type reductions has been suggested several years ego in \cite{ferap-khus}. This approach allowed one to find new classes of integrable partial differential equations \cite{ferap-khus}-\cite{fer-khus-tsar}, \cite{SokOdes}. An effective method of classification of fully discrete multidimensional equations using consistency
around a multidimensional cube is developed in \cite{ABS}. 

In the present article we propose a new classification algorithm suitable for integrable 2+1 dimensional lattices of the form
\begin{equation}\label{glattice}
\frac{\partial^2}{\partial x\partial y}u_n=f_n(u_n,u_{n,x},u_{n,y},u_{n\pm1},u_{n\pm1,x},u_{n\pm1,y},...
u_{n\pm k},u_{n\pm k,x},u_{n\pm k,y}),\, 
\end{equation}
where $-\infty<n<\infty$ and $u=u_n(x,y)$ is a function of three independent variables $x,y,n$. 

Our approach is based on the notion of characteristic vector fields introduced and applied to the classification problem of integrable 1+1 dimensional hyperbolic type PDE by E.Goursat \cite{Goursat}. In the modern context of integrability the concept of characteristic Lie rings was revived in \cite{LSSh}, where it was applied to Darboux integrable exponential type systems. Several attempts have been undertaken to adopt the characteristic Lie rings to the classification problem of S-integrable 1+1 dimensional continuous and discrete models \cite{ZhberMur}, \cite{HabGudkova1}, \cite{HabGudkova2}. 

We consider characteristic vector fields in essentially different situation, since the lattice (\ref{glattice}) actually is 2+1 dimensional. Note that multidimensional models do not admit local integrals and therefore Darboux integrability loses its sense in this case. Nevertheless one can assign two characteristic Lie rings $L_x$ and $L_y$ to any lattice of the form (\ref{glattice}). Indeed, operator $D_y=\frac{d}{dy}$ of the total derivative with respect to $y$ acting on the subset of dynamical variables $u_i,u_{i,x},u_{i,xx},...$ defines a vector field given as a formal series
\begin{equation}\label{2basicY}
Y=\sum_{i=-\infty}^{\infty}u_{i,y}\frac{\partial}{\partial{u_i}}+f_i\frac{\partial}{\partial{u_{i,x}}} +f_{i,x}\frac{\partial}{\partial{u_{i,xx}}}+f_{i,xx}\frac{\partial}{\partial{u_{i,xxx}}}+\cdots .
\end{equation}
Call it characteristic vector field in $y$-direction for the lattice (\ref{glattice}). 
Denote through \textbf{L}$_y$ the Lie ring generated by the operators $Y$, $X_i=\frac{\partial}{\partial{u_{i,y}}}$, 
where $i$ ranges in $(-\infty,\infty).$ Thus the characteristic Lie ring of an infinite lattice has an infinite set of generators therefore it is rather complicated to work with\footnote{In a slightly different way  characteristic Lie rings (algebras) for the special class of  exponential type 2+1 dimensional lattices were introduced  in \cite{Sh}.}. However it has a sequence of relatively simple subrings, which  can be used as indicators of integrability. For integrable lattices these subrings are of finite dimension.

In a similar way one can define the Lie ring \textbf{L}$_x$ generated by the operators $\tilde Y$, $\tilde X_i=\frac{\partial}{\partial{u_{i,x}}}$, where
\begin{equation}\label{2basictildeY}
\tilde Y=\sum_{i=-\infty}^{\infty}u_{i,x}\frac{\partial}{\partial{u_i}}+f_i\frac{\partial}{\partial{u_{i,y}}} +f_{i,y}\frac{\partial}{\partial{u_{i,yy}}}+f_{i,yy}\frac{\partial}{\partial{u_{i,yyy}}}+\cdots .
\end{equation}
Construct a sequence of the operators 
\begin{equation}\label{sequence}
X_i,\,Y_{1,i},\,Y_{2,i}, \,Y_{3,i},\ldots,
\end{equation}
where $i\in (-\infty,\infty)$ and $Y_{1,i}=[X_i,Y],$ $Y_{2,i}=[Y_{1,i},Y],$ $Y_{3,i}=[Y_{2,i},Y],\ldots\,Y_{k+1,i}=[Y_{k,i},Y], \ldots$.

Let \textbf{F} be a ring of locally analytic functions depending on a finite number of the dynamical variables $u_i$, $u_{i,x}$, $u_{i,y}$, $u_{i,xx}$, $u_{i,yy},\ldots .$ Fix a finite subset \textbf{A}$_y$ of the operators in (\ref{sequence}). Stress that \textbf{A}$_y$ may contain the operators $X_i$, $Y_{k,j}$ with different values of their indices, for example, one can choose \textbf{A}$_y$$=\left\{X_1; Y_{2,3}; Y_{4,1}\right\}$.  Denote through \textbf{L}$_{0,y}$ the set containing all of the operators in this subset, all possible commutators of these operators and linear combinations of the operators in \textbf{A}$_y$ and commutators with the coefficients from \textbf{F}. Actually \textbf{L}$_{0,y}$ has a structure of the left \textbf{F}-module (see, \cite{Bourbaki}). Recall that the left \textbf{F}-module \textbf{L}$_{0,y}$ is called finitely generated if there exist $Z_1$, $Z_2$, ..., $Z_m$ in \textbf{L}$_{0,y}$ such that for all $Z$ in \textbf{L}$_{0,y}$, there exist $r_1,$ $r_2,$ ..., $r_m$ in \textbf{F} with $Z = r_1Z_1 + r_2Z_2 + ... + r_m Z_m.$ The set $Z_1$, $Z_2$, ..., $Z_m$ is called a finite generating set of the module \textbf{L}$_{0,y}$. 
\begin{rem}\label{basis}
Finite generating set of \textbf{F}-module allows one to define the dimension of the module. Suppose that equation
$$r_1Z_1 + r_2Z_2 + ... + r_m Z_m=0$$
holds for the functions $r_1,$ $r_2,$ ..., $r_m$ analytic in a domain of many variable complex space. By reducing the domain if necessary one can get only two possible choices: either $r_1\neq 0$ everywhere in the domain or $r_1\equiv0$. In the former case we see that $Z_1$ is linearly expressed in terms of $Z_2$,... $Z_m$. Continuing this way we find a subset of linearly independent operators. Thus the module has a structure of a finite dimensional linear space.
\end{rem}

In a similar way one defines the left \textbf{F}-module \textbf{L}$_{0,x}$ in the $x$-direction, by choosing a finite subset \textbf{A}$_x$ of the  operator sequence, defined as follows 
$$\tilde X_i,\, \tilde Y_{1,i}=[\tilde X_i,\tilde Y],\, \tilde Y_{2,i}=[\tilde Y_{1,i},\tilde Y], \, \tilde Y_{3,i}=[\tilde Y_{2,i},\tilde Y],...$$

The following definition seems to be reasonable.
\begin{deff}\label{deff1}. A lattice of the form (\ref{glattice}) is called "integable" if  for any choice of the subsets \textbf{A}$_y$, \textbf{A}$_x$ the corresponding left \textbf{F}-modules \textbf{L}$_{0,y}$ and \textbf{L}$_{0,x}$ are finitely-generated.
\end{deff}

Theorems \ref{thmToda} and \ref{thmTodatype} below show that a large class of lattices known to be integrable are also "integrable" in the sense of Definition \ref{deff1}. Therefore Definition \ref{deff1} is regarded as an adequate formalization of the integrability property for the lattice (\ref{glattice}). Actually it can be used as a classification tool for integrable lattices.

Let us formulate some useful commutativity relations between generators of the rings \textbf{L}$_x$, \textbf{L}$_y$. For the simplicity concentrate on \textbf{L}$_y$, because all the properties of \textbf{L}$_x$ are exactly symmetrical.
\begin{lem}\label{lem1} Operators $X_i$, $Y$ acting on the variables $u_{i,y},u_i,u_{i,x},u_{i,xx},u_{i,xxx},...$ satisfy the relations
\begin{equation}\label{commrelations}
[D_x,X_i]=-X_i(f_i)X_i, \quad [D_x,Y]=-\sum_i Y(f_i)X_i.
\end{equation}
\end{lem}
Lemma can be proved by applying both sides of the relations to the dynamical variables listed.
\begin{lem}\label{lem2}
 Suppose that the vector field
 \begin{equation}
Z=\sum_j{a_j(1)\frac{\partial}{\partial{u}_{jx}}+a_j(2)\frac{\partial}{\partial{u}_{jxx}}+a_j(3)\frac{\partial}{\partial{u}_{jxxx}}+...}
 \end{equation}
 satisfies the equation $\left[D_x,Z\right]=0 $, then $Z=0$. 
\end{lem}
{\bf Proof}. Apply both sides of the equation $\left[D_x,Z\right]=0$ to the variables $u_j,u_{j,x},u_{j,xx},...$ and get an infinite system of equations
\begin{eqnarray}
&&a_j(1)=0,\nonumber\\
&&a_j(2)=D_x(a_j(1)),\nonumber\\
&&a_j(3)=D_x(a_j(2)),\nonumber\\
&&............  \nonumber\\
&&a_j(k+1)=D_x(a_j(k)),\nonumber\\
&&............  \nonumber
\end{eqnarray}
which immediately implies that all of the coefficients $a_j(k)$ vanish. Lemma is proved.

The article is organized as follows. In Section 2 we recall the definition of characteristic Lie rings for systems of hyperbolic type equations. It is used in Section 4 but also it explains the motivation of introducing similar objects in 2+1 dimensional case. In Section 3 we adopt our definitions to Volterra type lattices. In the last Section 4 we prove Theorem 2 and formulate Theorem 3 approving the integrability conjecture given in Definitions \ref{deff1}, \ref{deff2}.

\section{Characteristic Lie rings for systems of hyperbolic type PDE's}

Consider a system of hyperbolic type partial differential equations of the following form
\begin{equation}\label{hsystem}
\frac{\partial^2}{\partial x\partial y}u_i=g_i({\bf u},{\bf u_x}, {\bf u_y}),\,i=1,2,...N,
\end{equation}
where ${\bf u}=(u_1,u_2,...u_N)$, ${\bf u_x}=\frac{\partial}{\partial x}{\bf u}$, ${\bf u_y}=\frac{\partial}{\partial y}{\bf u}$. Define a set of standard dynamical variables $S=S_x\bigcup S_y$, where
\begin{equation}\label{dset}
S_x=\{{\bf u},{\bf u_x},{\bf u_{xx}},...\}, \quad S_y=\{{\bf u},{\bf u_y},{\bf u_{yy}},...\}.
\end{equation}
As usually we consider dynamical variables as independent ones.

Recall that a function $I$ depending on a finite number of the variables in $S_x$ is called $y$-integral if $D_yI=0$ by means of the  equation system (\ref{hsystem}). A set of $y$-integrals $I_{(1)}, I_{(2)},... I_{(N)}$ constitutes a complete set of integrals if none of these integrals is expressed in terms of the other ones and their total derivatives with respect to $y$. In a similar way the complete set of $x$-integrals for (\ref{hsystem}) is defined.

System (\ref{hsystem}) is called Darboux integrable if it admits complete sets of integrals in both $x$-
and $y$-directions.

According to the definition any $y$-integral solves the following equation $Y_{h}I:=D_yI({\bf u},{\bf u_x},{\bf u_{xx}},...)=0$, where due to the chain rule the characteristic vector field $Y_{h}$ is defined as follows
\begin{equation}\label{basicY}
Y_{h}=\sum_{i=1}^{N}u_{i,y}\frac{\partial}{\partial{u_i}}+g_i\frac{\partial}{\partial{u_{i,x}}} +g_{i,x}\frac{\partial}{\partial{u_{i,xx}}}+\cdots .
\end{equation}
Since the $y$-integral does not depend on the variables $u_{i,y}$ while the coefficients of $Y_{h}$ depend on them, we have to write in addition to the equation $Y_{h}I=0$ also equations $X_iI=0$ for $i=1,2,...N$, where $X_i=\frac{\partial}{\partial{u_{i,y}}}$.

Let \textbf{l}$_{h,y}$ be the set containing all of the operators $X_1,\,X_2\,...,X_N,\,Y_h$, all possible commutators of these operators and linear combinations of the operators and commutators with the variable coefficients in \textbf{F}. Then \textbf{l}$_{h,y}$ has a structure of the left \textbf{F}-module. Thus any $y$-integral is annulated by the elements of \textbf{l}$_{h,y}$. In a similar way the module \textbf{l}$_{h,x}$ is defined.

The following theorem establishes a correspondence between integrals and the characteristic Lie rings (see, for instance, survey  \cite{ZMHSh}).

\begin{thm}\label{char}
System (\ref{hsystem}) admits a complete set of $y$-integrals (or $x$-integrals) iff its characteristic \textbf{F}-module \textbf{l}$_{h,y}$ (respectively, characteristic \textbf{F}-module \textbf{l}$_{h,x}$) is finitely generated.
\end{thm}

\section{Volterra type lattices.}

In the case of the Volterra type lattices 
\begin{equation}\label{Volterra}
u_{n,y}=p(u_n,v_{n+1},v_n),\quad v_{n,x}=q(v_n,u_n,u_{n-1})
\end{equation}
our definitions are slightly changed. The characteristic vector fields in $y$-direction are defined as follows
\begin{equation}\label{vY}
Y_v=\sum_{i=-\infty}^{\infty}p_{i}\frac{\partial}{\partial{u_i}}+p_{i,x}\frac{\partial}{\partial{u_{i,x}}} +p_{i,xx}\frac{\partial}{\partial{u_{i,xx}}}+\cdots .
\end{equation}
and $X_{v,i}=\frac{\partial}{\partial{v_{i}}}.$ Now define the sequence 
\begin{equation}\label{sequencev}
X_{v,i},\, Y_{v,1,i}=[X_{v,i},Y_v],\, Y_{v,2,i}=[Y_{v,1,i},Y_v],\, Y_{v,3,i}=[Y_{v,2,i},Y_v],\, \ldots\,Y_{v,k+1,i}=[Y_{v,k,i},Y_v], \ldots ,
\end{equation}
where integer $i$ ranges in $(-\infty,\infty)$.
Having the operator sequence  we choose its finite subset \textbf{A}$_{v,y}$ and construct the characteristic module \textbf{L}$_{v,0,y}$.

Similarly we define the characteristic vector fields in $x$-direction
\begin{equation}\label{vtildeY}
\tilde Y_v=\sum_{i=-\infty}^{\infty}q_{i}\frac{\partial}{\partial{v_i}}+q_{i,y}\frac{\partial}{\partial{v_{i,y}}} +q_{i,yy}\frac{\partial}{\partial{v_{i,yy}}}+\cdots .
\end{equation}
\begin{equation}\label{vtildeXY}
\tilde X_{v,i}=\frac{\partial}{\partial{u_{i}}}.
\end{equation}
Afterwards we choose a finite set \textbf{A}$_{v,x}$ and construct the corresponding characteristic module \textbf{L}$_{v,0,x}$.

\begin{deff}\label{deff2}. A lattice of the form (\ref{Volterra}) is called "integable" if  for any choice of the subsets \textbf{A}$_{v,y}$, \textbf{A}$_{v,x}$ the corresponding modules \textbf{L}$_{v,0,y}$ and \textbf{L}$_{v,0,x}$ are finitely-generated.
\end{deff}
 
\section{Fundamental property of integrable Toda and Volterra type lattices.}

Sequences of characteristic modules introduced in the previous sections are approved to be an effective tool to justify integrability of a given lattice (see Definitions \ref{deff1}, \ref{deff2} above).  It is observed that for a large class of  integrable lattices of the forms (\ref{glattice}) and (\ref{Volterra}) the characteristic modules in both directions are finitely generated for any choice of the operator sets. Let us illustrate the statement with the example of Toda lattice \cite{M}:
\begin{equation}\label{Toda}
\frac{\partial^2}{\partial x\partial y}u_n=exp\{2u_n-u_{n-1}-u_{n+1}\},\, -\infty<n<\infty.
\end{equation}

\begin{thm}\label{thmToda} For any choice of the sets \textbf{A}$_{y}$, \textbf{A}$_{x}$ for the Toda lattice (\ref{Toda}) the corresponding modules \textbf{L}$_{0,y}$ and \textbf{L}$_{0,x}$ are finitely-generated.
\end{thm}

{\bf Proof}. Reduce the infinite lattice (\ref{Toda}) to a finite system of hyperbolic type PDE's by imposing the following cutting off boundary conditions $u_{-N-1}=0$ and $u_{N+1}=0$ which are known to preserve integrability of lattice (\ref{Toda}). As a result we find a finite system of hyperbolic type partial differential equations:
\begin{eqnarray}{}
\frac{\partial^2}{\partial x\partial y}u_{-N}&=&exp\{2u_{-N}-u_{-N+1}\},\nonumber\\
\frac{\partial^2}{\partial x\partial y}u_n&=&exp\{2u_n-u_{n-1}-u_{n+1}\},\, -N<n<N,\label{Tfinite}\\
\frac{\partial^2}{\partial x\partial y}u_N&=&exp\{2u_N-u_{N-1}\}\, \nonumber.
\end{eqnarray}
Consider the Lie ring generated by the operators $X_{-N}$, $X_{-N+1}$,...,$X_{N}$, $Y_{s}$ where $X_i=\frac{\partial}{\partial{u_{i,y}}}$, and 
\begin{equation}\label{NbasicY}
Y_{s}=\sum_{i=-N}^{N}u_{i,y}\frac{\partial}{\partial{u_i}}+g_i\frac{\partial}{\partial{u_{i,x}}} +g_{i,x}\frac{\partial}{\partial{u_{i,xx}}}+\cdots ,
\end{equation}

where 
\begin{eqnarray}{}
&&g_{-N}=exp\{2u_{-N}-u_{-N+1}\},\nonumber\\
&&g_i=exp\{2u_i-u_{i-1}-u_{i+1}\}, \, -N<i<N,\nonumber\\
&&g_N=exp\{2u_N-u_{N-1}\}\nonumber.
\end{eqnarray}

System (\ref{Tfinite}) is known to be Darboux integrable (see \cite{LSSh}), i.e. it admits complete sets of integrals in both characteristic directions. Hence due to the Theorem \ref{char} characteristic Lie rings \textbf{l}$_{x,N}$ and \textbf{l}$_{y,N}$ have a structure of finitely generated \textbf{F}-modules. Moreover in the case of the Toda lattice these modules are finite dimensional linear spaces, they admit finite basis (see Remark \ref{basis}). 

Turn back to the lattice (\ref{Toda}) which is a particular case of (\ref{glattice}). The following statement takes place.
\begin{lem}\label{N}
For any natural $k$ and any integer $i$ we have $Y_{k,i}\in $\textbf{l}$_{y,N}$ and $\tilde Y_{k,i}\in $\textbf{l}$_{x,N}$ for $N\geq |i|+k$.
\end{lem}

{\bf Proof of the Lemma \ref{N}}. By definition we have $X_i\in $\textbf{l}$_{y,N}$ for $|i|\leq N$. For (\ref{Toda}) and its reduction (\ref{Tfinite}) we have $Y_{1,i}=[X_i,Y]=\frac{\partial}{\partial u_i}$ and  $Y_{s,1,i}=[X_i,Y_s]=\frac{\partial}{\partial u_i}\in $\textbf{l}$_{y,N}$ hence $Y_{1,i}\in $\textbf{l}$_{y,N}$ for $|i|\leq N$.

Comparison of the operators $Y_{2,i}=[Y_{1,i},Y]$ and $Y_{s,2,i}=[Y_{s,1,i},Y_s]\in $\textbf{l}$_{y,N}$ given as formal series
\begin{equation}\label{Y2}
Y_{2,i}=\sum_{j=i-1}^{i+1}f_{j,u_i}\frac{\partial}{\partial{u_{i,x}}} +(f_{i,x})_{u_i}\frac{\partial}{\partial{u_{i,xx}}}+\cdots ,
\end{equation}
and 
\begin{equation}\label{Y2s}
Y_{s,2,i}=\sum_{j=i-1}^{i+1}g_{j,u_i}\frac{\partial}{\partial{u_{i,x}}} +(g_{i,x})_{u_i}\frac{\partial}{\partial{u_{i,xx}}}+\cdots 
\end{equation}
shows that $Y_{2,i}=Y_{s,2,i}$ for $|i|\leq N-2$ since $f_j=g_j$ for $|j|\leq N-1$. Therefore, $Y_{2,i}\in $\textbf{l}$_{y,N}$ for $|i|\leq N-2$. Continuing this way one can complete the  proof of the Lemma by induction proving  that $Y_{k,i}\in $\textbf{l}$_{y,N}$ for $|i|\leq N-k$, where $N>k$. It implies that $Y_{k,i}\in $\textbf{l}$_{y,N}$ for $N\geq |i|+k$. 

End of proof of the Theorem \ref{thmToda}. Choose a finite set \textbf{A}$_y$ of operators in (\ref{sequence}) corresponding to the Toda lattice. Due to Lemma \ref{N} there is a sufficiently large natural $N$ such that \textbf{A}$_y \subset $\textbf{l}$_{y,N}$.  As it was noted above \textbf{l}$_{y,N}$ is a finite dimensional linear space, therefore \textbf{L}$_{0,y}$ generated by the set \textbf{A}$_y$ is also a finite dimensional linear space. Thus it is a finitely generated \textbf{F}-module.
The theorem is proved.

In a similar way one can prove the following statement concerned to the lattices listed in \cite{ShY} as the integrable ones.
\begin{thm}\label{thmTodatype} 
For arbitrary choice of the sets \textbf{A}$_{y}$, \textbf{A}$_{x}$ for any of the lattices below the corresponding modules \textbf{L}$_{0,y}$ and \textbf{L}$_{0,x}$ are finitely-generated
\begin{eqnarray}
u_{nxy}=\exp (u_{n+1}-u_{n})-\exp (u_{n}-u_{n-1}),\label{2}\\
u_{nxy}=\exp (u_{n-1})-2\exp (u_{n})+\exp (u_{n-1}),\label{3}\\
u_{nxy}=u_{nx}(u_{n+1}-2u_{n}+u_{n-1}),\label{4}\\
u_{nxy}=u_{nx}[\exp (u_{n+1}-u_{n})-\exp (u_{n}-u_{n-1})],\label{5}\\
u_{ny}=u_n(v_{n+1}-v_n), \quad v_{nx}=u_n-u_{n-1},\label{6}\\
u_{ny}=u_n(v_{n+1}-v_n), \quad v_{nx}=v_n(u_n-u_{n-1}).\label{7}
\end{eqnarray}
\end{thm}

For the lattice (\ref{3}) the proof of the statement easily follows from the results in \cite{Sh}.

\section*{Conclusions}

The concept of characteristic Lie rings provides an effective tool to investigate Darboux integrable 1+1 dimensional nonlinear equations. To the best of our knowledge there is no any analogue of this notion for multidimensional equations. In this article an attempt is undertaken to fill up this gape, the concept of characteristic Lie rings is generalized to 2+1 dimensional lattices. A sequence of special type subsets is considered. It is shown that for the well-known integrable lattices all of these subsets are finitely generated modules. This fact can be used as a basis of a  classification criterion for integrable lattices. An intriguing question is about relation of Definition \ref{deff1} and the  method of hydrodynamic type reductions proposed in \cite{ferap-khus}. The matter is that the finitely generated modules mentioned above are closely connected with the infinite sequence of Darboux integrable reductions of the lattice.

\section*{Acknowledgments}

This work is partially supported by Russian Foundation for Basic Research (RFBR) grants $\#$ 11-01-97005-r-povolzhye-a and $\#$10-01-00088-a and by Federal Task Program "Scientific
and pedagogical staff of innovative Russia for 2009-2013" contract no. 2012-1.5-12-000-1003-011.

\end{document}